# Comprehensive Multimodal Deep Learning Survival Prediction Enabled by a Transformer Architecture: A Multicenter Study in Glioblastoma

Ahmed Gomaa, Yixing Huang, Amr Hagag, Charlotte Schmitter, Daniel Höfler, Thomas Weissmann, Katharina Breininger, Manuel Schmidt, Jenny Stritzelberger, Daniel Delev, Roland Coras, Arnd Dörfler, Oliver Schnell, Benjamin Frey, Udo S. Gaipl, Sabine Semrau, Christoph Bert, Rainer Fietkau, and Florian Putz

Department of Radiation Oncology, University Hospital Erlangen, Friedrich-Alexander-Universität Erlangen-Nürnberg, Erlangen, Germany (A.G., Y.H., A.H., C.S., D.H., T.W., B.F., U.S.G., S.S., C.B., R.F., F.P.); Comprehensive Cancer Center Erlangen-EMN (CCC ER-EMN), Erlangen, Germany (A.G., Y.H., A.H., C.S., D.H., T.W., M.S., J.S., D.D., R.C., A.D., O.S., B.F., U.S.G., S.S., C.B., R.F., F.P.); Department Artificial Intelligence in Biomedical Engineering, Friedrich-Alexander-Universität Erlangen-Nürnberg, Erlangen, Germany (K.B.); Institute of Neuroradiology, University Hospital Erlangen, Friedrich-Alexander-Universität Erlangen-Nürnberg, Erlangen, Germany (M.S., A.D.); Department of Neurology, University Hospital Erlangen, Friedrich-Alexander-Universität Erlangen-Nürnberg, Erlangen, Germany (J.S.), Department of Neurosurgery, University Hospital Erlangen, Friedrich-Alexander-Universität Erlangen-Nürnberg, Erlangen, Germany (D.D., O.S.), Institute for Neuropathology, University Hospital Erlangen, Friedrich-Alexander-Universität Erlangen-Nürnberg, Erlangen, Germany (R.C.),The Bavarian Cancer Research Center (BZKF), Erlangen, Germany (A.G., Y.H., M.S., F.P.), FAU Profile Center Immunomedicine (FAU I-MED), Friedrich-Alexander-Universität (FAU) Erlangen-Nürnberg, Schlossplatz 1, D-91054 Erlangen, Germany (U.G.)

**Corresponding Author**: Yixing Huang, Department of Radiation Oncology, University Hospital Erlangen, Universitätsstraße 27, 91054 Erlangen, Germany (yixing.huang@uk-erlangen.de)

## Abstract

**Background**: This research aims to improve glioblastoma survival prediction by integrating MR images, clinical and molecular-pathologic data in a transformer-based deep learning model, addressing data heterogeneity and performance generalizability.

**Method**: We propose and evaluate a transformer-based non-linear and non-proportional survival prediction model. The model employs self-supervised learning techniques to effectively encode the high-dimensional MRI input for integration with non-imaging data using cross-attention. To demonstrate model generalizability, the model is assessed with the time-dependent concordance index (Cdt) in two training setups using three independent public test sets: UPenn-GBM, UCSF-PDGM, and RHUH-GBM, each comprising 378, 366, and 36 cases, respectively.

**Results**: The proposed transformer model achieved promising performance for imaging as well as non-imaging data, effectively integrating both modalities for enhanced performance (UPenn-GBM test-set, imaging Cdt 0.645, multimodal Cdt 0.707) while outperforming state-of-the-art late-fusion 3D-CNN-based models. Consistent performance was observed across the three independent multicenter test sets with Cdt values of 0.707 (UPenn-GBM, internal test set), 0.672 (UCSF-PDGM, first external test set) and 0.618 (RHUH-GBM, second external test set). The model achieved significant discrimination between patients with favorable and unfavorable survival for all three datasets (logrank p $1.9 \times 10^{-8}$, $9.7 \times 10^{-3}$, and $1.2 \times 10^{-2}$). Comparable results were obtained in the second setup using UCSF-PDGM for training/internal testing and UPenn-GBM and RHUH-GBM for external testing (Cdt 0.670, 0.638 and 0.621).

**Conclusions**: The proposed transformer-based survival prediction model integrates complementary information from diverse input modalities, contributing to improved glioblastoma survival prediction compared to state-of-the-art methods. Consistent performance was observed across institutions supporting model generalizability.

## Key Points

1. Improved glioblastoma survival prediction by combining multi-parametric MR images and non-imaging data using a transformer-based deep learning model.
2. Consistent model performance demonstrated across 3 institutions.
3. Non-linear and non-proportional survival function prediction.

## Importance of the Study

This work aims to improve the survival prediction accuracy of glioblastoma through a transformer-based deep learning model. The proposed model integrates multi-parametric MR images, clinical data, and molecular-pathologic markers by employing a self-supervised MRI encoder module and cross-attention fusion for individual patient survival function prediction. Model performance and generalizability are rigorously evaluated across public datasets from three institutions. The proposed transformer model consistently outperformed unimodal methods and enhanced glioblastoma survival prediction accuracy compared to state-of-the-art late-fusion 3D-CNN-based methods as well as conventional Cox regression models. Program code and data are publicly available to support reproducibility and implementation. By providing more accurate survival predictions in a context of inter-institutional data heterogeneity, this research could potentially contribute to the advancement of personalized medicine of individuals affected by glioblastoma.

## Introduction

Predicting the survival time of glioblastoma patients provides essential information for patients, clinicians, and healthcare systems and has always been a high interest of glioma research.[1] Improving survival prediction holds key importance for achieving more informed clinical decision making and personalized care. Prognosis of individual glioblastoma patients is highly heterogeneous and includes both extreme short- and long-term survival rendering optimal decision making in daily clinical practice challenging.[2] Moreover, an increasing amount of prognostic factors from a variety of data sources is known to contain complementary prognostic information that needs to be harnessed and integrated to achieve optimal survival prediction accuracy.[3-6] Prognostic information is contained in histologic morphology[3,6] as well as molecular pathologic data,[3,7,8] clinical and treatment-related parameters[9,10] like performance status, age, and extent of resection, and in imaging data[6,11,12] like MRI, and PET datasets. Integrating image information into a more comprehensive prediction of an individual patient's prognosis has been challenging for conventional methods of prognostication that have been designed for low-dimensional parameters only.[13] The high dimensionality of imaging data and its substantial inter- and intra-institutional heterogeneity so far have hindered wide-spread clinical availability of prognostic models that combine both imaging and non-imaging data in glioblastoma patients.

Multiple approaches have been pursued to tap into the prognostic information embedded in imaging data. First, human experts can extract well-described low-dimensional parameters from imaging data like tumor maximum cross-sectional area,[14] presence of the T2/FLAIR mismatch sign,[15] or ADC ratios.[16] While this approach is relatively robust and successful, it is suitable for a limited set of imaging features only. Second, a large number of predefined radiomic imaging features can be automatically extracted from an image segment using computer-aided analysis. In the Radiomics paradigm, this feature extraction step is followed by feature reduction and conventional machine learning models, which also allows for integration of imaging features with non-imaging parameters.[17] However, the predefined radiomic features are characterized by a relatively low-level of abstraction making them susceptible to vulnerabilities, intra- and inter-institutional imaging data heterogeneity (particularly for MRI), overfitting and consecutively reduced model generalizability.[18] In addition, the reliance on predefined segmentations can introduce additional biases as well as observer-dependence and may miss important imaging information outside the analyzed image compartment.[18,19] Third, deep learning models, like convolutional neural networks (CNNs) and vision transformers (ViT) learn the meaningful imaging features during the training process themselves, not only greatly increasing the variety of learnable features, but also allowing for a much higher level of abstraction. Therefore, deep learning image analysis methods have demonstrated superior performance and model generalizability.[20-22] Notably, given the relative scarcity of medical imaging data, a common two-stage approach has become prominent. In this approach, a CNN encoder is initially trained on large datasets, often including 2D non-medical images, such as ImageNet.[21,23-25] Subsequently, this previously trained network is fine-tuned or used as a fixed encoder medical image. [26,27]

Recently, significant progress in glioma prognosis prediction has been made by CNN-based deep learning models. Li et al. described a 2D-CNN DeepRisk network architecture to predict a risk score from 32 2D MRI slices extracted from multimodal MRI sequences around a preexisting tumor segmentation that was able to stratify patients into three risk groups.[28] More recently, Lee et al. proposed a 3D-CNN-based discrete-time survival model for predicting a single deep learning prognostic index from multi-modality whole-brain MRI datasets.[29] In this work, Lee et al. proposed a Deep Learning-based Prognostic Index (DPI), which quantifies spatial features from MRIs that correlate with patient survival. Late fusion was subsequently used in order integrate the DPI with clinical and molecular-genetic variables in a conventional Cox Proportional-Hazards (CoxPH) model. The state-of-the-art late-fusion 3D-CNN + CoxPH model proposed by Lee et al. will be subsequently referred to as 3D-CNN+CoxPH throughout this manuscript. Using a public 1085 training/validation and two private test sets, Lee et al. could show an improvement in conventional c-index from 0.774 to 0.783 and from 0.748 to 0.766 by adding the imaging-derived deep learning prognostic index to the non-imaging parameters.[30]

Conventional CoxPH models face limitations in survival prediction,[31] as they primarily calculate a relative risk function, allowing straightforward hazard ratio calculation but complicating the prediction of an individual patient's survival function in clinical practice. Moreover, CoxPH models rely on the assumption of proportional hazards, demanding constant hazard ratios between any two individuals throughout the observation period, and they lack the ability to account for non-linear associations between input covariates and event hazards for patients.[31] Furthermore, traditional regression models are suboptimal for handling the increasing number of input parameters from novel molecular-pathologic methods.[32] To address these constraints, adopting deep learning models for the entire survival prediction task becomes a viable strategy. Deep learning models exhibit the capability to capture complex, non-linear interactions among input covariates, enhancing predictive capabilities for survival outcomes.[33-35] The integration of multiple data streams, such as clinical records, medical imaging, and molecular-pathologic data, through deep learning has the potential to reveal hidden associations and interdependencies that might be obscured in unimodal models.[36-38] Transformer-based survival prediction models, with their cross-attention mechanism, are well-suited for efficiently integrating multimodal input data, offering advantages over CNN-based models.[23,39,40]

In the present work, we therefore propose and evaluate a transformer architecture as an exclusively deep learning-based strategy for survival prediction in glioblastoma patients. Our approach integrates prognostic information from multi-parametric MR input along with clinical and pathologic information such as gender, age, resection status, and O[6]-methylguanine-DNA methyltransferase (MGMT) promoter methylation status. In this approach, a transformer model is used for encoding the high-dimensional whole brain MR imaging data, integration of encoded imaging and non-imaging information, as well as for individual patient survival function prediction.

We make the following contributions:

1. A fully deep learning-based method is proposed for survival function prediction using a transformer architecture. Novel technical contributions include a self-supervised Vision Transformer for effective encoding of the local and global information from MR imaging input with reduced risk of over-fitting in the survival prediction task. In addition, to the best of our knowledge, this is the first work to use cross-attention to capture the interactions between clinical and molecular-pathologic data with multimodal MR images for individual patient survival function prediction.

2. We show that the proposed transformer architecture outperforms state-of-the-art late-fusion 3D-CNN based approaches as well as conventional CoxPH regression models for multimodal survival prediction in glioblastoma patients.

3. Consistent model performance is demonstrated for public datasets from three institutions in two training setups, indicating model generalizability.

## Materials and Methods

### Neural Network Architecture

We present a two-step multimodal learning framework, illustrated in Figure 1. In the initial step, a ViT, acting as an encoder, employs self-supervised learning to acquire a clinically-relevant encoding from multimodal whole-volume MR input datasets. In the second step, the encoded MRI inputs undergo integration with the encoded clinical and molecular-pathologic input using cross-attention. To prevent overfitting, the trained ViT with frozen parameters is utilized in this step. The outputs of the cross-attention, capturing interactions between modalities, are then passed to 2 fully-connected layers for final survival prediction using attention pooling. The succeeding paragraphs delve deeper into the intricacies of this process, offering a more comprehensive understanding of our proposed framework.

**Feature Extraction**: To bridge the feature space gap between input modalities, we employ specialized encoders for each modality. Clinical and molecular-pathologic data, with their structured nature, are encoded using a trainable fully-connected module trained end-to-end. However, the more complex volumetric MR input images undergo self-supervised pre-training for effective encoding of the brain's anatomical information (Figure 1 A). Inspired by the self-supervised training of Swin UNETR,[41] the trained imaging encoder model can learn how to generate clinically relevant representations from the high dimensional imaging input, which can be efficiently fused with non-imaging clinical data for various medical downstream tasks. The self-supervised pre-training process is formulated as a multi-objective loss function wherein the ViT-based encoder learns two proxy tasks, namely contrastive learning and context restoration. The selection of a transformer-based encoder instead of its conventional CNN-based counterpart is attributed to its capability of capturing global context and long-range dependencies in high-dimensional inputs, owing to its attention mechanism.[42,43] During pre-training, the multimodal MR volumes (T2, T2-FLAIR, T1 pre- and post-contrast) for each subject are stacked in the channel dimension. Subsequently, the volumes are augmented twice with random patch swapping and cutout, resulting in two views for each subject. The augmented views are then employed in the proxy tasks: context restoration, and contrastive learning. Context restoration aids the encoder in grasping the structural intricacies and anatomical context of different brain regions. In parallel, self-supervised contrastive learning endeavors to learn representations that encapsulate discriminative and pertinent information from the data. For further insights into the implementation details of this self-supervised learning methodology, additional information is provided in the Supplementary Material.

**Cross-attention**: For the fusion approach, we adapt the cross-attention methodology, introduced by Lu et al. for visio-linguistic representations.[44] Cross-attention is a type of attention mechanism used in machine learning models, particularly in natural language processing (NLP) and computer vision tasks,[44,45] to facilitate interactions between two different inputs or sequences. In this fusion method, each modality is mutually reinforced with relevant information from the other via multi-head attention.[46] The process of cross-attention enables the model to selectively focus on relevant parts of one sequence while processing another sequence, capturing the relationships between the two. Cross-attention can be extended to effectively model the interactions between two different modalities by attending one modality while processing the other.[45] Given two representation vectors from two modalities, α and β, we can utilize cross-attention to generate a new, enriched representation of α based on β. The updated, enriched representation vector, denoted as $\hat{\alpha}$, is given by the following equation:

$$\hat{\alpha} = \text{softmax}\left(\frac{Q_\alpha K_\beta^\top}{\sqrt{d_k}}\right) V_\beta$$

$$= \text{softmax}\left(\frac{\alpha W_{Q_\alpha} W_{K_\beta} \beta^\top}{\sqrt{d_k}}\right) \beta W_{V_\beta}.$$

In this equation, $W_{Q_\alpha}$, $W_{K_\beta}$, and $W_{V_\beta}$ represent learnable weights that are used to generate the queries $Q_\alpha$, keys $K_\beta$, and values $V_\beta$, while $d_k$ is a scaling factor. In our implementation, the cross-attention mechanism is employed twice. In each instance, one modality is projected onto the query vector $Q$, while the other is projected onto the key $K$ and value $V$ vectors. Consequently, this process yields enriched representations for each modality, drawing information from the other.

**Fusion and Prediction**: Upon obtaining the updated representations for each modality through the preceding cross-attention layers, a concatenation of both vectors occurs in the sequence dimension. Subsequently, this concatenated set of high-level representations is channeled into the final attention layer, where it assumes the roles of key and value vectors alongside with the clinical data projection as the query vector. The rationale for this approach stems from an empirical observation indicating that the non-imaging modality carries a higher degree of prognostically relevant information in comparison to the imaging modality. Finally, attention pooling is performed on the resultant vector, which is then passed to two fully connected layers for output prediction.

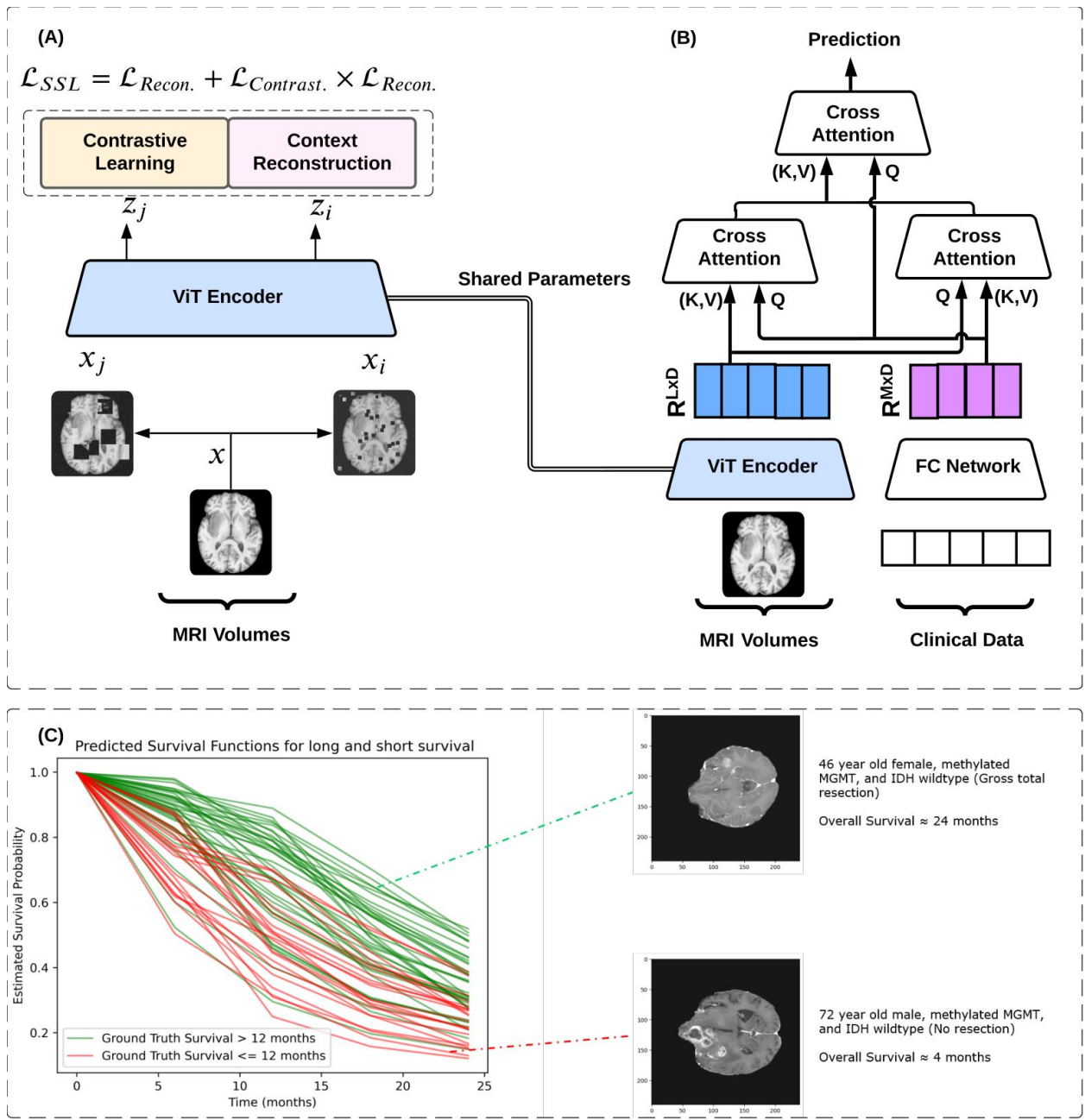

*Figure 1 The workflow of the proposed transformer architecture: (A) The self-supervised pre-training module, which is optimized to generate low-dimensional representation from the high-dimensional MRI Images. (B) The downstream attention-based multimodal survival prediction task. (C) Predicted survival curves from the downstream model with two example cases characterized by favorable and unfavorable survival.*

## Discrete-time Survival Prediction

Discrete-time survival models are statistical tools employed to analyze the likelihood of experiencing a specific event, within predefined time intervals These models break the follow-up period into discrete intervals to estimate event probabilities within each interval.[47] In light of this, the proposed model stands as a discrete-time survival model, employing a deep neural network for its parameterization following the *DeepHit* setup introduced by Lee et al. in [30]. In this setup, the model is designed to provide a comprehensive understanding of the joint distribution of the first event time and competing, or single, events while effectively handling right-censored data. The input for this setup encompasses three primary elements for each instance or patient. First, observed covariates ($x$) represent the patient's unique characteristics and relevant factors that may influence their survival outcomes. Second, the model incorporates the time elapsed ($s$) denoting the specific time at which the event of interest or censoring occurred. Third, it accounts for event labels ($k$ out of $K$ events), indicating the type of event that transpired, whether it be a particular adverse clinical event or censoring ($\emptyset$) when the event has not been observed. An essential feature of this setup's architectural design lies in its output layer, which employs a single softmax layer on the output vector $Y = [y_{1,1}, \cdots, y_{1,T_{max}}, \cdots, y_{K,T_1}, \cdots, y_{K,T_{max}}]$, to ensure that the model learns the probability $\hat{P}(s, k \mid x)$ of experiencing an event $k$ at time $s$.

The model's training loss function has two key parts. The first part penalizes the misordering of event pairs, enhancing accurate event sequencing. The second part, the log-likelihood term, captures the joint distribution of event times, accounting for right-censored data. A detailed analysis of the loss function components is included in the Supplementary Material.

Integral to understanding this model's operation is the estimation of the Cumulative Incidence Function (CIF), denoted as $\hat{F}_k(s^{(i)} \mid x^{(i)})$. This function quantifies the likelihood of event $k$ happening on or before a specific time point, and it is approximated by $\sum_{m=0}^{s} y_{k,m}$. The CIF plays an integral role in deriving the survival function of each patient, which can be obtained as:

$$S(t) = 1 - \hat{F}_k(s^{(i)} \mid x^{(i)})$$

In this work, $K$ is set to 1, as the single event of death is considered for prediction of the patient-specific overall-survival function. Ultimately, the model predicts the survival probability at five specific time points predetermined by the percentiles of the survival time within the training cohort. Subsequently, to establish survival probabilities on a monthly basis, a linear interpolation is applied, facilitating the estimation of survival probabilities for intermediate time intervals between these predefined points. Thus, the final output of the model is the prediction of an individual patient's survival function at a monthly resolution throughout the follow-up period (Figure 1 B).

## Patient Data and Multimodal Datasets

In this multicenter retrospective study, four independent public multimodal glioblastoma datasets are used to allow for optimal reproducibility and implementation. The three multicenter cohorts include the University of Pennsylvania Glioblastoma (UPenn) dataset (n = 378),[48,49] the University of California San Francisco Preoperative Diffuse Glioma MRI (UCSF) dataset (n = 366),[49,50] and the Rio Hortega University Hospital Glioblastoma (RHUH) dataset (n = 36).[49,51] In addition to the imaging data, these datasets encompass clinical and molecular-pathologic information, such as gender, age, resection status, and MGMT promoter methylation status. These features were chosen because they are well-documented and available in all three datasets we used, namely UPenn-GBM, UCSF-PDGM, and RHUH-GBN. Unfortunately, certain potentially valuable features like the Karnofsky Performance Status (KPS) were not available in the UCSF-PDGM dataset, which necessitated their exclusion to maintain uniformity in our feature set across all datasets.

The characteristics for each dataset are summarized in Table 1. For the self-supervised training of the image encoder module, n = 1251 multimodal MRI studies from the BraTS 2021[52-54] dataset (imaging data only) are utilized, excluding 298 duplicate cases that were also part of the UCSF dataset. Approval has been obtained from the ethics committee at Erlangen University Hospital to conduct this study. Additionally, the studies conducted in the used public datasets have been approved by their respective ethics committee.[48,50,51]

*Table 1 Clinical and molecular-pathologic variables in the three multicenter datasets.*

| Variable | UCSF (n = 366) | UPenn (n = 378) | RHUH (n = 36) |
|---|---|---|---|
| Age (years), mean ± SD | 61.8 ± 12.03 | 63.8 ± 11.7 | 63.7 ± 9.01 |
| Median overall survival, days | 359 | 373 | 341.5 |
| Sex | | | |
|    Male, n (%) | 218 (59.6%) | 232 (61.4%) | 24 (66.6%) |
|    Female, n (%) | 148 (40.4%) | 146 (38.6%) | 12 (33.4%) |
| Extent of Resection | | | |
|    GTR [> 90% resection], n (%) | 213 (58.2%) | 214 (56.6%) | 36 (100%) |
|    NTR [< 90% resection], n (%) | 0 (0%) | 146 (38.6%) | 0 (0%) |
|    NA, n (%) | 153 (41.8%) | 18 (4.8%) | 0 (0%) |
| MGMT Status | | | |
|    Unmethylated, n (%) | 103 (28.1%) | 139 (36.8%) | 0 (0%) |
|    Methylated, n (%) | 246 (67.2%) | 72 (19.0%) | 0 (0%) |
|    Not available, n (%) | 17 (4.7%) | 167 (44.2%) | 36 (100%) |
| KPS | | | |
|    ≥ 80, n (%) | - | 52 (13.7%) | 20 (55.5%) |
|    < 80, n (%) | - | 14 (3.7%) | 16 (44.5%) |
|    NA, n (%) | 366 (100%) | 312 (82.6%) | 0 (0%) |

## Data Preprocessing

The four main morphological MRI modalities in glioblastoma T2, T2-FLAIR, and T1 pre- and post-contrast are used. Preprocessing steps include spatial normalization, bias field correction and stripping of any residual skull parts. We stack the 4 morphological volumes in the channel dimension and perform histogram standardization and z-normalization. For the clinical data, the non-categorical covariates are normalized to have values between -1 and 1. Categorical covariates, on the other hand, were converted to one-hot encoded vectors.

## Evaluation and Statistical Analysis

The proposed model is comprehensively evaluated using statistical metrics and survival analysis techniques. The time-dependent concordance index (Ctd), an extension to Harrell's C-index,[55] assesses predictive accuracy over time,[56] and the integrated Brier score (IBS) measures differences between observed and predicted survival probabilities. Kaplan-Meier survival analyses with Logrank tests and Schoenfeld criteria for the hazard proportionality assumption are conducted. Results show that, except for age and resection status, all features meet the criterion (p ≥ 0.05). The transformer model is evaluated in two scenarios: one with 70% UPenn patients were used for training, while the remaining cases were split evenly for validation and internal testing, and another with roles of UPenn and UCSF datasets switched. Consistent censoring and survival time among the dataset splits are considered in both setups, and test sets remain unused until the final evaluations to ensure unbiased results. Detailed results and tests for the proportionality assumption are in the Supplementary Material.

## Results

### Model Training Using the Multimodal University of Pennsylvania Dataset (Setup 1)

In the first setup, model training, validation and internal testing was performed on the University of Pennsylvania Glioblastoma (UPenn) dataset, while the University of California San Francisco (UCSF) dataset and the Rio Hortega University Hospital (RHUH) dataset served as two independent external test sets (Table 2).

**Patient Characteristics**:

The training set included 264 patients, characterized by a mean age of 63.6 ± 11.6 years, and a male predominance (159 males). The median overall survival within this set was 381 days, with an interquartile range spanning 182-558 days. The validation set, consisting of 57 patients, exhibited a mean age of 64.3 ± 12.1 years, and a male distribution of 38 individuals. Within this subset, the median survival was observed to be 360 days, with an interquartile range extending from 167 to 550 days. Similarly, the internal test set, also comprising 57 patients, featured a mean age of 63.9 ± 11.8 years, with 34 males. The median survival in this set was recorded as 370 days, and the interquartile range was reported to be 182-553 days.

**Prognostic Accuracy - Clinical and Molecular-Pathologic Data Only**:

To apply the proposed transformer model to non-imaging data only, cross-attention was exchanged with self-attention. The transformer model was compared against the conventional Cox Proportional-Hazards model (CoxPH) as well as the two deep learning survival models DeepSurv[33] and DeepHit[30] as state-of-the-art methods for non-imaging data. CoxPH displayed competitive performance when trained on UPenn data, with Ctd-values of 0.667 ± 0.009 (UPenn, internal test set), 0.642 ± 0.016 (UCSF, first external test), and 0.552 ± 0.022 (RHUH, second external test). However, DeepSurv[33] and DeepHit[30] showed greater variations in performance across institutions (Table 2). The proposed transformer architecture exhibited relatively consistent performance across all institutions with Ctd values of 0.667 ± 0.011 (UPenn), 0.669 ± 0.021 (UCSF), and 0.613 ± 0.021 (RHUH) outperforming the state-of-the-art methods for the test sets.

**Prognostic Accuracy – Imaging Data Only:**

For the end-to-end Vision Transformer model (ViT), without self-supervised pre-training, Ctd values were 0.622 ± 0.012 (UPenn), 0.578 ± 0.019 (UCSF), and 0.603 ± 0.021 (RHUH). The proposed self-supervised framework outperformed ViT for all test sets, achieving Ctd values of 0.645 ± 0.014 (UPenn), 0.615 ± 0.020 (UCSF), and 0.609 ± 0.021 (RHUH).

**Prognostic Accuracy – Combined Imaging and Non-Imaging Data:**

Combining non-imaging and imaging data improved the performance of the proposed transformer model. The model consistently achieved higher Ctd-indices than for any single modality: 0.707 ± 0.012 (UPenn), 0.672 ± 0.023 (UCSF), and 0.618 ± 0.014 (RHUH). The recently proposed 3D-CNN-based late-fusion model by Lee et al.[29](3D-CNN+CoxPH) served as state-of-the-art method. This 3D-CNN based method uses a deep learning model to extract a prognostic score from the MRI data that is subsequently combined with the clinical and pathologic parameters using a conventional CoxPH model. The proposed transformer architecture showed higher Ctd-values than the 3D-CNN-based late-fusion model for all external test sets (0.671 ± 0.010 for UPenn, 0.659 ± 0.021 for UCSF, and 0.607 ± 0.020 for RHUH).

Using the median survival time of the training sets (12.0 months) as a dichotomization threshold, the proposed multimodal model achieved significant discrimination between patients with favorable and unfavorable survival for all test sets (p-values $1.9 \times 10^{-8}$, $9.7 \times 10^{-3}$, and $1.2 \times 10^{-2}$ for UPenn, UCSF, and RHUH, respectively). Figure 2(A) shows the Kaplan-Meier survival plots for each dataset.

## Model Training Using the Multimodal University of California San Francisco Dataset (Setup 2)

In the second setup, the roles of the UPenn and UCSF datasets were exchanged to further evaluate model generalization in the context of changing training datasets. The UCSF dataset served for training, validation, and internal testing, whereas external testing was performed on the UPENN and RHUH datasets (Table 2).

**Patient Characteristics:**

In this setup, the training set comprised 256 patients, with a mean age of 62.0 ± 11.7 years (152 males). The median overall survival for this cohort equaled 357 days, with interquartile range spanning 169 to 616 days. Furthermore, the validation set encompassed 55 patients with a mean age of 61.8 ± 14.2 years and 27 males included. The median survival within this subset was calculated as 334 days, and the interquartile range extended from 156 to 715 days. Similarly, the test set, consisting of 55 patients, demonstrated a mean age of 60.5 ± 11.3 years, with 39 males. The median survival in this set mirrored the training set at 365 days, while the interquartile range spanned from 195 to 632 days.

**Prognostic Accuracy - Clinical and Molecular-Pathologic Data Only:**

In a parallel fashion to the initial scenario, the CoxPH model exhibited relatively favorable performance with a noteworthy Ctd of 0.647 ± 0.014 (UCSF), 0.633 ± 0.017 (UPenn), and 0.598 ± 0.020 (RHUH). Furthermore, akin to the first scenario, both DeepSurv and DeepHit exhibited varying degrees of performance consistency across the test datasets. Conversely, the proposed transformer model consistently demonstrated competitive predictive capabilities, yielding C-indices of 0.664 ± 0.011 (UCSF), 0.638 ± 0.020 (UPenn), and 0.614 ± 0.019 (RHUH).

**Prognostic Accuracy – Imaging Data Only:**

Echoing the findings from first configuration, the proposed self-supervised transformer framework consistently outperformed the end-to-end ViT across all clinical centers, demonstrating Ctd-values of 0.610 ± 0.016 (UCSF), 0.617 ± 0.019 (UPenn), and 0.609 ± 0.018 (RHUH).

**Prognostic Accuracy – Combined Imaging and Non-Imaging Data:**

Similar to the first configuration, the proposed model consistently delivered improved Ctd-indices of 0.670 ± 0.011 (UCSF), 0.637 ± 0.021 (UPenn), and 0.621 ± 0.019 (RHUH) when combining imaging and non-imaging data. In addition, the proposed transformer architecture showed higher Ctd-values than the 3D-CNN-based late-fusion model for all test sets (0.682 ± 0.009 for UCSF, 0.569 ± 0.023 for UPENN and 0.596 ± 0.014 for RHUH).

Finally, similar to the first configuration, the multimodal transformer model achieved significant discrimination between patients with favorable and unfavorable outcome (p-values $5.6 \times 10^{-6}$, $3.1 \times 10^{-2}$, and $7.9 \times 10^{-3}$ for UCSF, UPenn, and RHUH, respectively, Figure 2(B)).

Table 2 Summary of results for all training setups and input modalities. The evaluation metric and the 95% confidence interval are reported as sample mean ± margin. The best Cdt values for each input modality and setup are highlighted in bold.

| | Data | Model | Metric | UPenn (Internal) | UCSF (External 1) | RHUH (External 2) |
|---|---|---|---|---|---|---|
| **Setup 1 (Model Training on UPenn Dataset)** | Clinical and Molecular-Pathologic Data | CoxPH[31] | Cdt | 0.667 ± 0.009 | 0.642 ± 0.016 | 0.552 ± 0.022 |
| | | | IBS | 0.093 ± 0.008 | 0.092 ± 0.011 | 0.145 ± 0.007 |
| | | DeepSurv[33] | Cdt | 0.644 ± 0.011 | 0.613 ± 0.024 | 0.570 ± 0.021 |
| | | | IBS | 0.129 ± 0.004 | 0.167 ± 0.014 | 0.145 ± 0.005 |
| | | DeepHit[30] | Cdt | 0.641 ± 0.007 | 0.567 ± 0.023 | 0.603 ± 0.019 |
| | | | IBS | 0.242 ± 0.004 | 0.203 ± 0.010 | 0.241 ± 0.007 |
| | | **Proposed Transformer** | Cdt | **0.667 ± 0.011** | **0.669 ± 0.021** | **0.613 ± 0.021** |
| | | | IBS | 0.111 ± 0.006 | 0.162 ± 0.012 | 0.172 ± 0.006 |
| | Imaging Data | ViT[40] end-to-end | Cdt | 0.622 ± 0.012 | 0.578 ± 0.019 | 0.602 ± 0.023 |
| | | | IBS | 0.173 ± 0.006 | 0.208 ± 0.014 | 0.192 ± 0.009 |
| | | **Proposed self-supervised** | Cdt | **0.645 ± 0.014** | **0.615 ± 0.020** | **0.609 ± 0.021** |
| | | | IBS | 0.171 ± 0.007 | 0.197 ± 0.019 | 0.186 ± 0.012 |
| | Combined | 3D-CNN+CoxPH[29] | Cdt | 0.671 ± 0.010 | 0.659 ± 0.021 | 0.607 ± 0.020 |
| | | | IBS | 0.081 ± 0.006 | 0.151 ± 0.009 | 0.141 ± 0.008 |
| | | **Proposed Transformer** | Cdt | **0.707 ± 0.012** | **0.672 ± 0.023** | **0.618 ± 0.014** |
| | | | IBS | 0.157 ± 0.003 | 0.179 ± 0.021 | 0.180 ± 0.006 |

| | Data | Model | Metric | UCSF (Internal) | UPenn (External 1) | RHUH (External 2) |
|---|---|---|---|---|---|---|
| **Setup 2 (Model Training on UCSF Dataset)** | Clinical and Molecular-Pathologic Data | CoxPH[31] | Cdt | 0.647 ± 0.014 | **0.638 ± 0.017** | 0.598 ± 0.020 |
| | | | IBS | 0.095 ± 0.005 | 0.077 ± 0.019 | 0.142 ± 0.007 |
| | | DeepSurv[33] | Cdt | 0.611 ± 0.010 | 0.599 ± 0.022 | 0.523 ± 0.023 |
| | | | IBS | 0.161 ± 0.009 | 0.109 ± 0.014 | 0.177 ± 0.006 |
| | | DeepHit[30] | Cdt | 0.645 ± 0.013 | 0.562 ± 0.019 | 0.539 ± 0.021 |
| | | | IBS | 0.112 ± 0.008 | 0.184 ± 0.010 | 0.217 ± 0.013 |
| | | **Proposed Transformer** | Cdt | **0.664 ± 0.011** | 0.638 ± 0.020 | **0.614 ± 0.019** |
| | | | IBS | 0.183 ± 0.008 | 0.174 ± 0.013 | 0.201 ± 0.011 |
| | Imaging Data | ViT[40] end-to-end | Cdt | 0.602 ± 0.014 | 0.603 ± 0.021 | 0.591 ± 0.022 |
| | | | IBS | 0.187 ± 0.010 | 0.188 ± 0.011 | 0.194 ± 0.009 |
| | | **Proposed self-supervised** | Cdt | **0.610 ± 0.016** | **0.617 ± 0.019** | **0.609 ± 0.018** |
| | | | IBS | 0.186 ± 0.009 | 0.185 ± 0.014 | 0.189 ± 0.012 |
| | Combined | 3D-CNN+CoxPH[29] | Cdt | 0.650 ± 0.010 | 0.528 ± 0.018 | 0.587 ± 0.020 |
| | | | IBS | 0.115 ± 0.007 | 0.096 ± 0.009 | 0.161 ± 0.007 |
| | | **Proposed Transformer** | Cdt | **0.670 ± 0.009** | **0.638 ± 0.022** | **0.621 ± 0.019** |
| | | | IBS | 0.172 ± 0.006 | 0.191 ± 0.018 | 0.184 ± 0.009 |

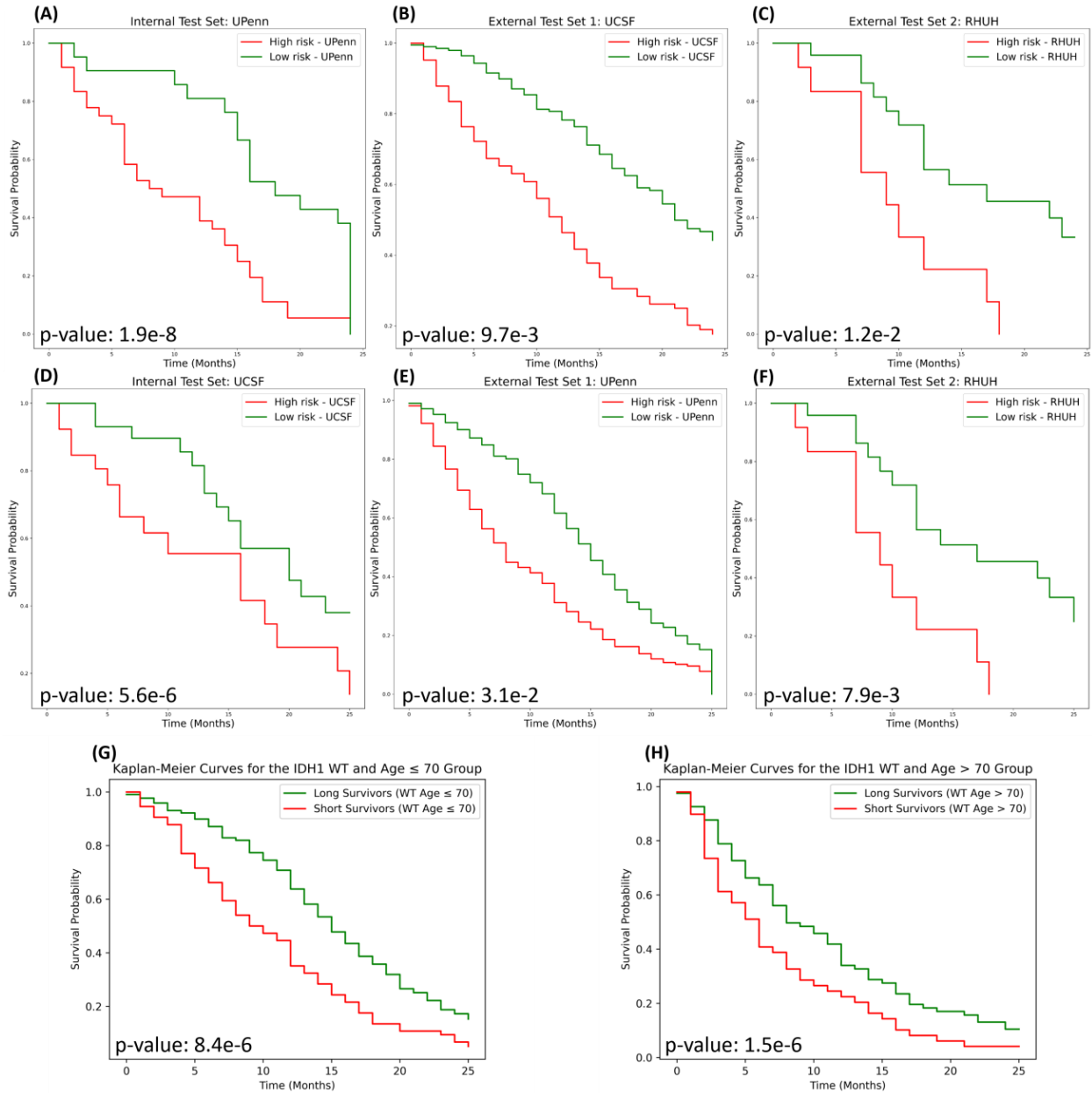

*Figure 2 Kaplan-Meier plots of observed survival separated into two prognostic groups by the model predictions for the internal and external test sets of both investigated configurations. The top row (subplots A, B, and C) illustrates the results for the first setup (training/validation/internal testing: UPENN dataset, external testing: UCSF and RHUH cohorts). Whereas the second setup (training/validation/internal testing: UCSF dataset, external testing: UPENN and RHUH cohorts) is shown in the bottom row (subplots D, E, and F). Only the results for the internal test and external test sets are shown. The bottom-most row demonstrates the prognostic stratification by the deep learning model predictions within distinct age cohorts for the first setup: (G) patients ≤70 years old and (H) patients > 70 years old.*

## Deep Learning Model Explainability

**Feature Representation**

Examination of the feature representation of the patient data and survival functions were conducted. We implemented the t-distributed Stochastic Neighbor Embedding (t-SNE) methodology[57] to visualize the learned representations derived from the amalgamated multimodal data. The resulting visualization, as depicted in Figure 3(A-B), portrays the learned representations of the test datasets projected onto a two-dimensional space.

Notably, a clear clustering pattern is evident: patients with mutated MGMT methylation status are predominantly grouped in one region of the embedding space, while those with wild-type MGMT methylation status occupy the opposite region. Similarly, patients who underwent a gross total resection of more than 90% cluster distinctly from those who did not. These spatial arrangements are consistent with the survival curves shown in Figure 3(C-D), which indicate that tumors with MGMT methylation mutations and higher resection rates are associated with improved prognoses. The coherence between the spatial distribution of patient representations and the corresponding survival outcomes highlights the model's ability to differentiate and link key prognostic factors within the learned representations of patient data.

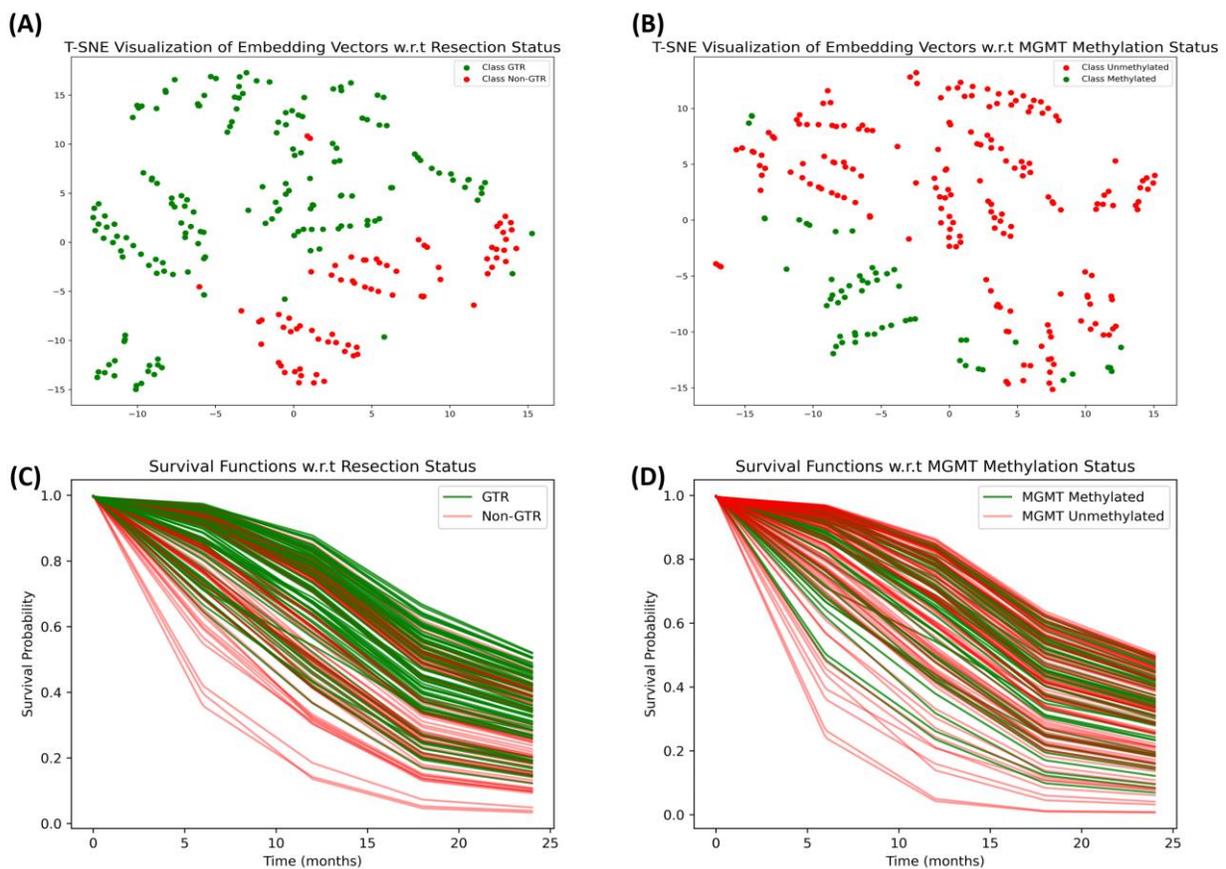

*Figure 3 The top row shows the t-SNE visualization of down-sampled patient data representations, depicting clear clustering based on gross total resection status and MGMT methylation. Regions in the embedding space distinctly separate patients with methylated MGMT status and those with gross total resection > 90% from their counterparts, correlating with improved survival outcome curves as shown in the bottom row.*

**Ablation Study**

To investigate the effect of individual components of the proposed deep learning architecture, we conducted a series of ablation experiments. The detailed results on the validation set of the UCSF dataset are summarized in Table 3. Firstly, when examining the self-supervised MRI encoder module, the experimental results show that regularizing the reconstruction loss with the contrastive loss improves performance of the downstream survival prediction. Using only the reconstruction loss for the image encoder resulted in a noticeable drop in the overall performance, indicating the necessity of the regularized loss. The empirical results show that regularizing the reconstruction loss with the contrastive loss led to the best performance. Subsequently, we examine the effectiveness of the proposed self-supervised MRI encoder module in comparison to a 2D-CNN ResNet[58,59] encoder either pretrained on ImageNet or trained in an end-to-end fashion. The findings demonstrate that the self-supervised encoder yielded superior performance.

Table 3 The ablation study results on the proposed mode showcasing the impact of the model's components.

| Model | Metric | UCSF |
| --- | --- | --- |
| Reconstruction Loss + Contrastive × Reconstruction Loss | Ctd | 0.662 |
|  | IBS | 0.171 |
| Reconstruction + Contrastive Loss | Ctd | 0.647 |
|  | IBS | 0.165 |
| Contrastive Loss | Ctd | 0.639 |
|  | IBS | 0.192 |
| Reconstruction Loss | Ctd | 0.623 |
|  | IBS | 0.188 |
| ViT Encoder (End-to-End) | Ctd | 0.632 |
|  | IBS | 0.210 |
| ResNet Encoder (ImageNet) | Ctd | 0.613 |
|  | IBS | 0.198 |

## Subgroup Analysis

We assessed the model predictions from setup 1 for differentiating favorable from unfavorable prognosis in the important subgroups of IDH1 wildtype glioblastoma patients ≤ 70 and > 70 years of age to evaluate the model predictions for treatment selection. Within the cohort of patients under 70 years, median survival time for the predicted short survivor group was observed to be 8.7 months, while the long survivor group demonstrated a median survival time of 14.8 months (log-rank p-value = $8.4 \times 10^{-6}$, n = 261). In the cohort aged 70 years or older, the median survival time for the predicted short survivor group was 5.6 months, while the long survivor group exhibited a median survival time of 8.1 months (p-value = $1.5 \times 10^{-2}$, n = 117).

## Discussion

Prognosis prediction has been a long-standing interest of glioma research.[60] Histologic morphology[3,6], molecular-pathologic factors[3,7,8], imaging features[6,15], clinical and treatment-related factors[5,9,10] have all been shown to predict the course of the disease and assist in decision making in glioblastoma patients. Important molecular-pathologic factors in glioma have been incorporated in the recent WHO classification[3] and selected clinical prognostic factors like age and performance status inform treatment algorithms in current neuro-oncologic guidelines.[61-63] However, for conventional methods, integration of the plethora of known prognostic factors into an improved comprehensive prediction has been challenging, especially considering incorporating high-dimensional imaging data together with structured clinical and molecular-pathologic parameters.[13] Novel deep learning models could be a key for such a comprehensive prognosis prediction integrating all available multimodal data streams for an individual patient, that could allow for more informed decision-making and treatment selection than selected single parameters are able to.

In this work, involving 780 patients from three datasets, we developed a transformer-based deep learning model, which integrates the prognostically relevant information from imaging, clinical and molecular-pathologic data for improved survival prediction. The model employs a specialized imaging encoder, trained in a self-supervised setup, to generate clinically relevant representation vectors from high-dimensional MRI input. These imaging representations are then fused with clinical data using the cross-attention mechanism. Based on the results presented in Table 2 the proposed model shows promising signs of improved performance when handling both uni- and multimodal inputs, outperforming DeepHit, conventional CoxPH regression, and its deep learning counterpart DeepSurv in clinical data analysis. In the case of imaging data, when compared against a ViT model trained end-to-end, the proposed model demonstrates better performance, suggesting that the transformer architecture has learned prognostically relevant representations in the self-supervised learning phase (Ctd of 0.662 for the proposed approach versus 0.632 for the end-to-end model). This drop in the performance for the end-to-end trained variant can be attributed to the increased variance of the model. A further advantage of the self-supervised training strategy is that unlabeled MR imaging data can be employed, which is much more abundant in neuro-oncology than annotated datasets. Conversely, the common two-stage approach of pretraining an MR encoder on non-medical photographs from the ImageNet database yielded subpar results compared to our model (Ctd of 0.613). This disparity in performance may be explained by the mismatch between MR data and the non-medical photographs from ImageNet, which can lead to the generation of clinically irrelevant representation vectors by the encoder. Notably, using multiple modalities with the proposed transformer architecture allowed for improvement in prognostic performance indicating that the model has learned how to combine complementary information from the different input modalities. Moreover, the proposed transformer model featuring a cross-attention fusion method showed improved performance over the late-fusion 3D-CNN- and CoxPH-based method introduced by Lee et al.[29] This can be explained by the late fusion's inability to capture interactions between the modalities before the final aggregation, which could prevent providing richer representations of the inputs. Finally, the proposed comprehensive transformer architecture has demonstrated consistent performance across external test sets for both training setups, suggesting the potential for generalizability.

The developed model addresses the non-linear and non-proportional aspects of survival predictions observed in glioblastoma. Traditional models like CoxPH assume linear relationships and proportional hazards, which may not capture the intricate dynamics in glioblastoma. The Schoenfeld test confirmed a significant departure from the proportional hazards assumption ($p < 0.05$) in the present multicenter cohorts for both *age* and *resection status* supporting the need for models capable of capturing complex interactions and non-linear relationships in neuro-oncology. The proposed transformer model offers potential for more granular and personalized survival predictions, overcoming limitations of traditional linear models and enhancing clinical decision-making for individual patients.

The subgroup analysis in younger and elderly glioblastoma patients aims to evaluate the potential application of the comprehensive transformer model in guiding treatment decisions, akin to the established use of the Karnofsky Performance Status (KPS) in current neurooncologic guidelines.[61,63] While direct comparisons with KPS weren't feasible in the available datasets because of a lack of continuous KPS data, the findings hint at the model's capacity to stratify survival outcomes within specific age brackets for glioblastoma patients. These insights encourage further investigation of transformer-based survival models for complementing established clinical parameters like KPS in guiding personalized treatment strategies within larger clinical trial cohorts.

This work has some limitations. First, this study does not account for missing modalities (either clinical or imaging) and assumes availability of the T2, T2-FLAIR, and T1 pre- and post-contrast modalities for all patients. Second, diffusion- and perfusion-weighted MRI and other advanced imaging techniques have the potential to further improve prognostic accuracy but were not available for the present analysis. Third, additional input modalities that could have further improved prognostic accuracy include microscopic imaging data like Whole Slide Images and advanced molecular-genetic analyses including genomic and methylation profiling could not be incorporated. Finally, while deep learning-based survival prediction has multiple advantages including precise individual patient survival function prediction, integration of imaging and non-imaging data and larger prognostic accuracy, it lacks the straight-forward interpretability of conventional models like CoxPH.

In this work, we introduce a transformer-based model for predicting survival in glioblastoma patients, combining multimodal MR images, clinical, and molecular-pathologic data. This model tackles data heterogeneity and computational challenges through self-supervised learning and attention mechanisms. It demonstrated consistent performance across various test sets, highlighting its generalizability. The use of multimodal MR and clinical data has shown to potentially enhance glioblastoma prognosis accuracy, potentially improving individualized treatment plans, such as optimal radiotherapy selection. Future directions include applying the MRI encoder for tasks like genetic biomarker status prediction or distinguishing between pseudo-progression and actual progression post-radiotherapy. The model could also be adapted for predicting outcomes for other brain tumor types, offering predictions on overall survival, local control, and distant-metastasis-free survival.

# Conflict of Interest

The authors declare no conflict of interest in this work.

# Funding

This work was partially funded by a grant of the Bavarian Cancer Research Center (BZKF).

# Authorship statement

The study was designed and conceived by FP, YH, and AG. AG wrote the program code, conducted the experimental work. AH contributed to the code. AG, YH and FP wrote the original draft. CB, KB, RF, AD, TW, DH and MS reviewed and edited the script. All authors have read and approved the script.